# Magnetotransport properties of (Ga,Mn)As investigated at low temperature and high magnetic field

T. Omiya[1], F. Matsukura[1], T. Dietl[1*], Y. Ohno[1], T. Sakon[2], M. Motokawa[2], and H. Ohno[1]

[1]Laboratory for Electronic Intelligent Systems, Research Institute of Electrical Communication, Tohoku University, Sendai 980-8577, Japan

[2]Institute for Materials Research, Tohoku University, Sendai 980-5877, Japan

**Abstract**

Magnetotransport properties of ferromagnetic semiconductor (Ga,Mn)As have been investigated. Measurements at low temperature (50 mK) and high magnetic field (≤ 27 T) have been employed in order to determine the hole concentration $p$ = 3.5x$10^{20}$ $cm^{-3}$ of a metallic $(Ga_{0.947}Mn_{0.053})As$ layer. The analysis of the temperature and magnetic field dependencies of the resistivity in the paramagnetic region was performed with the use of the above value of $p$, which gave the magnitude of p-d exchange energy $|N_0\beta|$ ~ 1.5 eV.

Corresponding author:

Hideo Ohno

Laboratory for Electronic Intelligent Systems, Research Institute of Electrical Communication, Tohoku University, Katahira 2-1-1, Aoba-ku, Sendai 980-8577, Japan, fax/phone +81-22-217-5553, e-mail: ohno@riec.tohoku.ac.jp

## 1. Introduction

Since the successful growth of the ferromagnetic III-V compound semiconductor (Ga,Mn)As [1], studies of its magnetotransport properties have been carried out [2-4]. It has been reported that (Ga,Mn)As shows p-type conduction with high hole concentrations ($p$ = $10^{18}$ – $10^{20}$ $cm^{-3}$) due to the substitution of divalent Mn for Ga in the host zinc-blende structure. This high concentration of the holes is believed to result in the hole-induced ferromagnetic order at low temperatures (<110 K). It is obvious that in order to understand further the origin of the ferromagnetism, an accurate determination of $p$ is necessary. The determination of $p$ from magnetotransport measurements is not simple, however, due to the existence of the anomalous Hall effect even up to room temperature and the negative magnetoresistance at low temperatures. Here, we report on the determination of hole concentration from the magnetotransport studies carried out at low temperature and high magnetic fields, where the anomalous Hall effect is almost completely saturated. We use, then, this value of $p$, and analyze the temperature and magnetic field dependencies of the resistivity in the paramagnetic region in terms of carrier scattering by thermodynamic fluctuations of magnetization.

## 2. Experimental

Depending on Mn content $x$, the temperature dependence of (Ga,Mn)As shows a metallic or insulating behavior. In this work, in order to avoid complications arising from localization [4], we focus on the behavior of metallic (Ga,Mn)As with Mn content $x$ = 0.053. A 200-nm thick (Ga,Mn)As layer with $x$ = 0.053 studied in this work was grown by molecular beam epitaxy onto a $(Al_{0.9}Ga_{0.1})As$ buffer layer residing on semi-insulating GaAs (100) substrate. The details of the growth procedure were described elsewhere [1,5].

Magnetotransport measurements were performed in the magnetic field $B$ perpendicular to the sample plane. Hall resistance $R_{Hall}$ and sheet resistance $R_{sheet}$, were measured simultaneously using Hall bar geometry. The measurement at temperature $T$ = 50 mK with $B$ up to 27 T was carried out in a hybrid-magnet system equipped with a dilution refrigerator. The temperature dependence of $R_{\mathrm{Hall}}$ and $R_{\mathrm{sheet}}$ in the range from 1.5 to 300 K with $B$ up to 10 T was measured by using a standard setup.

## 3. Results and discussion

Figure 1 shows the results of the magnetotransport measurements at 50 mK. $R_{\mathrm{Hall}}$ of magnetic materials can be expressed by the sum of the ordinary Hall resistance and the anomalous Hall resistance that is proportional to the perpendicular component of the magnetization $M$,

$$R_{\mathrm{Hall}} = R_0 B/d + R_s M/d, \qquad (1)$$

where $R_0$ is the ordinary Hall coefficient, $R_s$ the anomalous Hall coefficient, and $d$ thickness of the conducting layer. $R_s$ is proportional to either $R_{\mathrm{sheet}}$ or $R_{\mathrm{sheet}}^2$, depending on the mechanism of the anomalous Hall effect -- skew or side-jump scattering, respectively. As shown in Fig. 1(a), $R_{\mathrm{Hall}}$ shows a steep rise in the low field region (< 0.5 T) due to the rotation of magnetization from the in-plane orientation (magnetic easy plane) to the direction of $\boldsymbol{B}$, then drops slightly (< 8 T), and finally increases monotonically in the high field region (> 8 T). In the highest $B$ regime, the anomalous Hall resistance saturates because of the saturation of both magnetization and magnetoresistance (Fig. 1(b)). The remaining linear slope reflects the ordinary Hall term, from which one can determine the hole concentration unambiguously. We obtained $p$ = 3.5x10$^{20}$ cm$^{-3}$ from the slope of $R_{\mathrm{Hall}}$ versus $B$ curve above 22 T (the inset to Fig. 1), which is by a factor of 2.5 higher than the value determined when using the data taken at 5-7 T at 10 K ($p$ = 1.4x10$^{20}$ cm$^{-3}$). The present value of $p$ corresponds to 30 % of the nominal Mn content (~1.2x10$^{21}$ cm$^{-3}$), which suggests that there is a compensation of Mn acceptors by deep donors, most probably As antisite, a high concentration of which is known to be present in GaAs grown at low-temperature.

We hereafter use the hole concentration determined from the measurements shown in Fig.

1 in order to analyze the temperature and magnetic field dependencies of the magnetotransport properties in the high-temperature paramagnetic region. In this temperature range, we can neglect strong spin correlation, magnetic anisotropy, and localization effects seen at low temperatures. In the present analysis, we assumed that skew-scattering ($R_s \propto \rho$) controls the magnitude of the anomalous Hall effect, but the choice of skew scattering or side-jump scattering gives only a small effect (20 % error at most) on the final results.

As shown in Fig. 2, the temperature dependence of the inverse of paramagnetic susceptibility $\chi$, obtained from the magnetotransport properties around $B$ = 0 using eq. (1) is well described by the Curie-Weiss law with the Curie temperature $\theta$ = 105 K. Our result demonstrate that one can determine $\theta$ of (Ga,Mn)As from transport data alone.

The temperature dependence of the resistivity ($\rho = dR_{\mathrm{sheet}}$) of metallic (Ga,Mn)As shows a peak around ferromagnetic transition temperature $T_c$. Negative magnetoresistance is observed in the same temperature range ($\rho$ decreases by 20 % from 0 to 7 T at 100 K). The magnitude of negative magnetoresistance and the temperature dependence of $\rho$ suggest that spin disorder scattering by thermodynamic fluctuations of the magnetic spins is involved. A peak around $T_c$ can be interpreted as critical scattering by packets of the magnetic spins with a ferromagnetic short-range order characterized by the correlation length comparable to the wavelength of the carriers at the Fermi level. Within this interpretation the negative magnetoresistance is caused by the reduction of scattering associated with spin-alignment. The corresponding contribution to the resistivity is given by [6],

$$\rho_s = 2\pi^2 \frac{k_{\mathrm{F}}}{pe^2} \frac{m^2\beta^2}{h^3} \frac{k_{\mathrm{B}}T}{g^2\mu_B^2} \left(2\chi_\perp(T,B) + \chi_{//}(T,B)\right), \tag{2}$$

where $k_F$ is the Fermi wave vector, which is determined from $p$ assuming a spherical Fermi surface; $m$ is the hole effective mass taken here as $0.5m_0$ ($m_0$: the mass of free electron); $\beta$ the exchange integral of the interaction between the holes and the magnetic spins; $k_B$ the Boltzmann constant; $h$

the Planck constant; $g$ is the Landé factor of the Mn spins ($g$ = 2 is assumed), and $\mu_B$ the Bohr magneton. $\chi_\perp$ and $\chi_{//}$ are the transverse and longitudinal magnetic susceptibilities, respectively, which are determined from the magnetotransport data by the use of eq. (1); $\chi_\perp = M/B$, $\chi_{//} = \partial M/\partial B$. This formula, derived from the fluctuation-dissipation theorem, takes into account the presence of correlation between neighboring spins, $<\boldsymbol{S}_i\boldsymbol{S}_j> \neq <\boldsymbol{S}_i^2>\delta_{ij}$.

As shown in Fig. 3, the temperature dependence of $\rho$ at $B$ = 0 far above $T_c$ can be well reproduced by eq. (2) using $\beta$ and a background resistivitity as fitting parameters. The deviation near $T_c$ may be explained by noting that the long-range nature of the carrier-mediated magnetic interactions reduces the magnitude of $\chi$ for non-zero wave vectors, an effect disregarded in eq. (2). From the fit we obtained $\beta$ (p-d exchange), which is usually expressed in terms of $|N_0\beta|$ ($N_0$ is the density of the cation sites), as 1.7 eV.

The magnetic field dependence of $\rho$ at high temperatures is also well reproduced by scattering from the fluctuations of the magnetic spins, as shown in Fig. 4. Again, we used eq. (2) to fit the data at $T \geq 200$ K, and obtained $|N_0\beta|$ as 1.3 eV. In the previous analysis of the dependence of $\rho$ on $B$, we adobted an expression, which assumed non-correlated spin disorder. This leads us to obtain the upper limit of $|N_0\beta|$ which was 3.3 eV [3]. Actually, according to eq. (2), the new value of the hole concentration $p$ would lead to even greater value of $|N_0\beta|$. The difference between such a value of $|N_0\beta|$ and the present one thus reflects the important contribution of the ferromagnetic spin correlation to the magnitude of magnetoresistance.

Typically $|N_0\beta|$ is around 1 eV for II-VI based diluted magnetic semiconductors. According to magnetooptics studies [7, 8] on (Ga,Mn)As, $N_0\beta < 0$ ($> 0$) for $x = 0.005$ ($x = 0.00047$). At the same time $|N_0\beta|$ = 0.6 eV was determined by resonant tunneling spectroscopy [9], while $N_0\beta = -1.2$ eV resulted from the photoemission measurements [10]. According to the first principle theoretical calculation $N_0\beta < 0$ [11]. In order to determine $N_0\beta$ from optical measurements, one has to take into account the spin-splitting-induced redistribution of the holes between the spin subbands, the Moss-Burnstein shift caused by the high hole concentration [12], and the localization effect which enhances the exchange between carrier and magnetic spins through the carrier-carrier interaction [13]. The transport properties may also be influenced by such many-body effects.

## 4. Summary

We have presented the results of the magnetotransport measurements for $(Ga_{0.947}Mn_{0.053})As$. From the data at low temperature and in the high magnetic fields, we determined hole concentration as $3.5 \times 10^{20}$ cm$^{-3}$. The analysis of the temperature and magnetic field dependencies of the resistivity leads to the magnitude of the p-d exchange energy $|N_0\beta| \approx 1.5$ eV, assuming that the value of the hole effective mass is $m_h^* = 0.5m_0$.

## 5. Acknowledgments

The measurement at low-temperature and in high-magnetic fields was performed at the High Field Laboratory for Superconducting Materials, Institute for Materials Research, Tohoku University. This work was partly supported by the “Research for the Future” Program (# JSPS-RFTF97P00202) from JSPS and by a Grant-in-Aid on the Priority Area “Spin Controlled Semiconductor Nanostructures” (# 09244103) from the Ministry of Education, Japan.

## References

* Permanent address: Institute of Physics, Polish Academy of Sciences, PL-02668 Warszawa, Poland.

Figure Captions

Fig. 1 Magnetotransport properties of $(Ga_{0.947}Mn_{0.053})As$ at 50 mK in a high magnetic filed. (a) Hall resistance; as shown in the inset Hall resistance is the linear function of the magnetic field in the high field region. (b) Sheet resistance; negative magnetoresistance tends to saturate in the high field region.

Fig. 2 Temperature dependence of the inverse of the magnetic susceptibility which is determined from the magnetotransport measurements. Solid line shows the fitting using the Curie-Weiss law, which gives the Curie-Weiss temperature of 105 K.

Fig. 3 Temperature dependence of the resistivity in the high-temperature paramagnetic region. Solid squares and open circles in this figure show experimental data and the fitting using eq. (2)., respectively.

Fig. 4 Magnetic field dependence of the resistivity in the high-temperature paramagnetic region. Solid line shows the fitting using eq. (2).

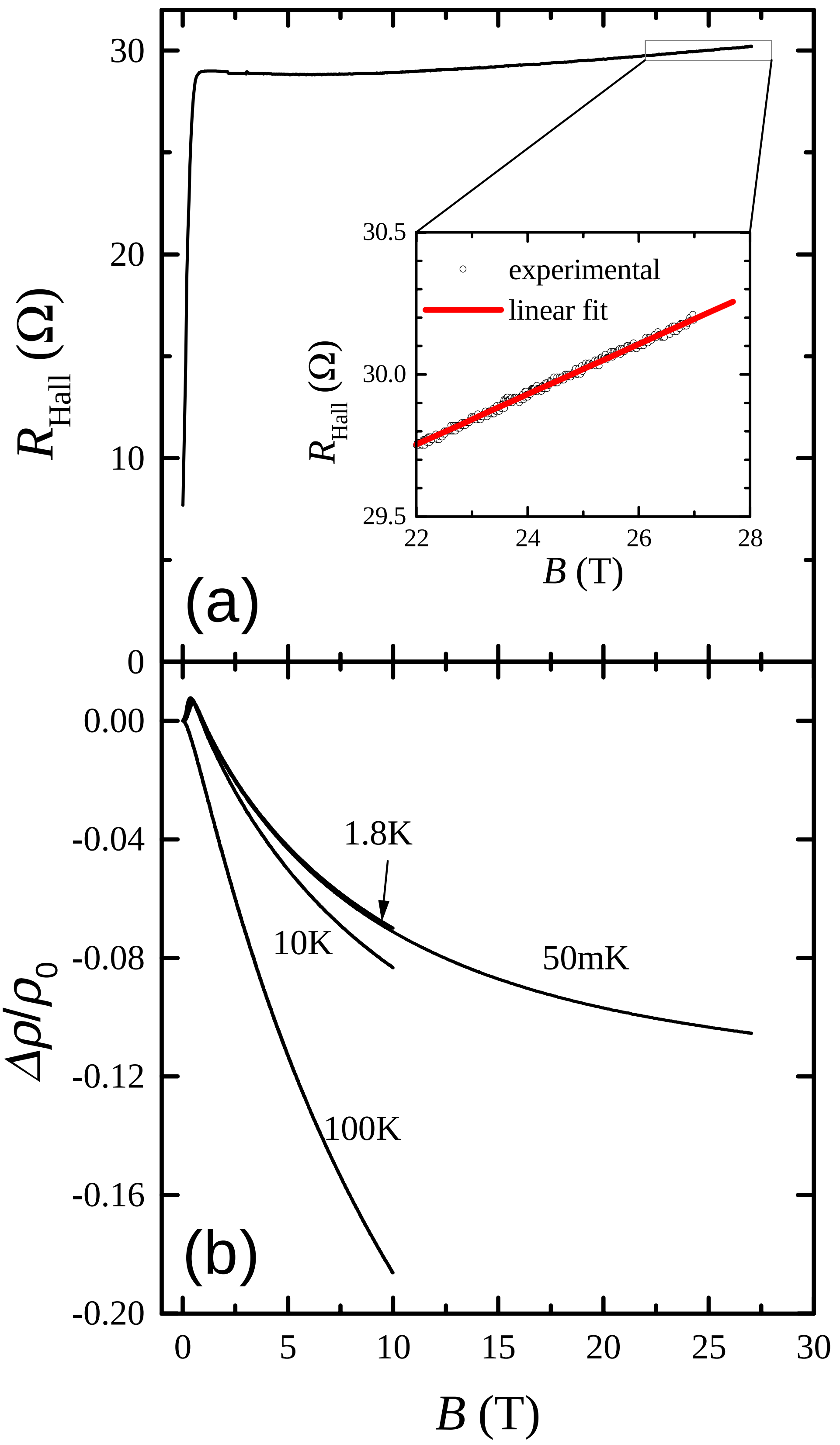

Fig. 1

Omiya *et al.*

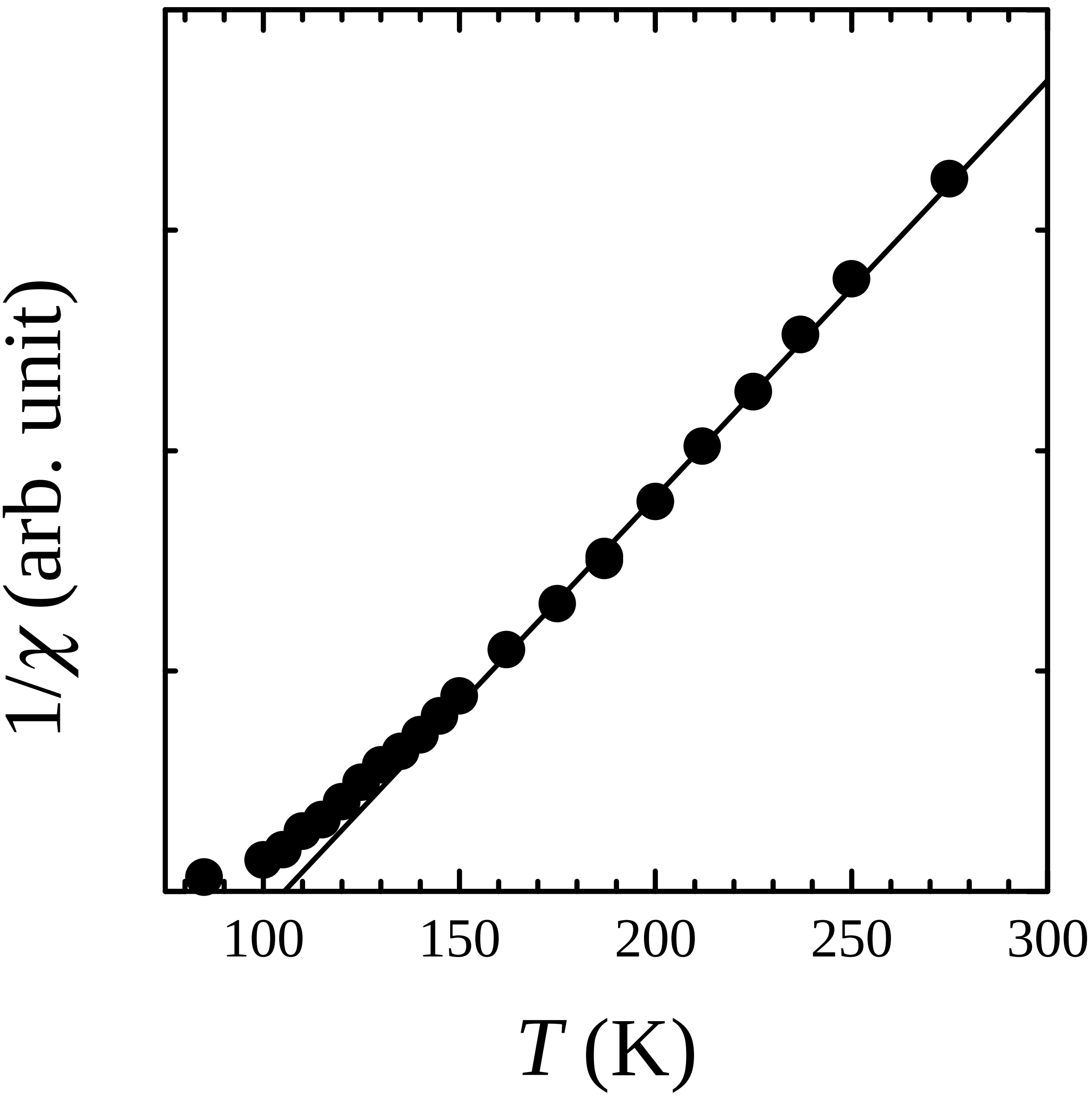

Fig. 2

Omiya *et al.*

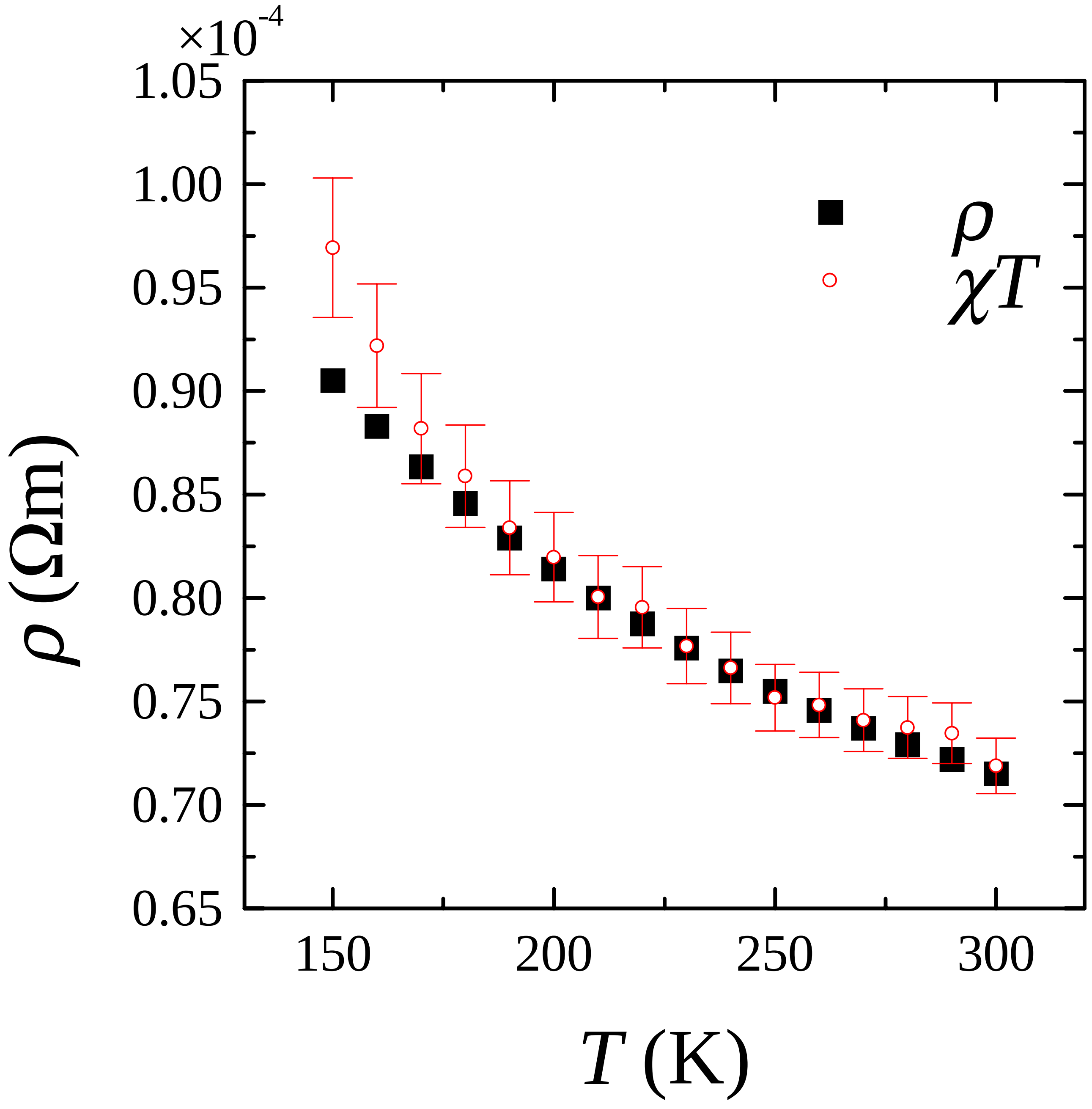

Fig. 3

Omiya *et al.*

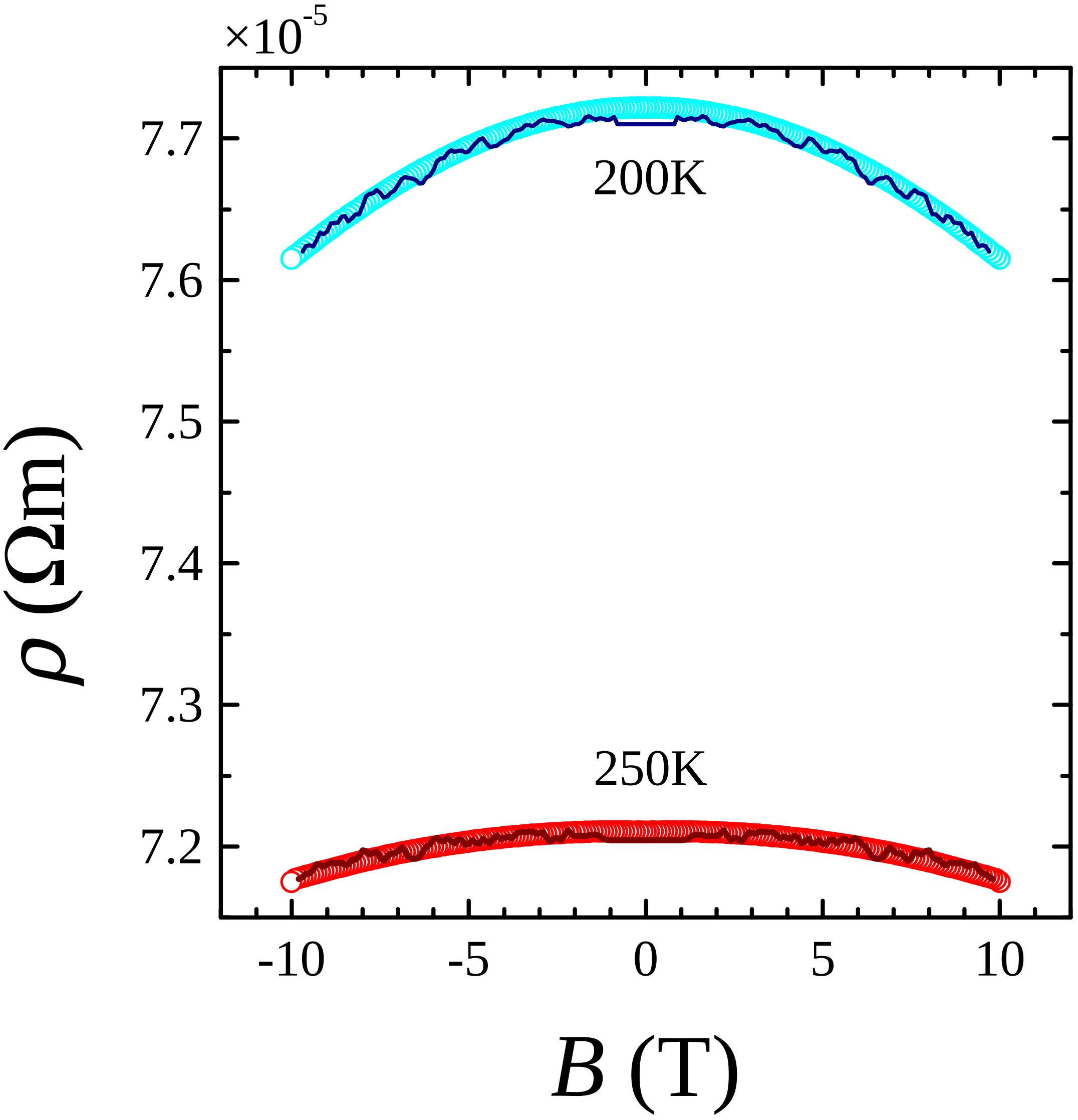

Fig. 4

Omiya *et al.*